\documentclass[]{aastex}

\usepackage{amsmath}
\usepackage{amssymb}
\usepackage{graphics}
\usepackage{rotating}

\newcommand{\hcn}{H$^{13}$CN}
\newcommand{\hnc}{HN$^{13}$C}
\newcommand{\aand}{$\&$ }

\newcommand{\kms}{km~s$^{-1}$}
\newcommand{\deltav}{$\delta v$}

\shorttitle{The inflow signature toward different evolutionary phases of massive star formation}
\shortauthors{Jin et al.}

\begin{document}

\title{The inflow signature toward different evolutionary phases of massive star formation}

\author{Mihwa Jin$^1$,  Jeong-Eun Lee$^1$, Kee-Tae Kim$^2$, Neal J. Evans II$^3$} 
\affil{
       $^1$ School of Space Research, Kyung Hee University, 1732, Deogyeong-daero, Giheung-gu, Yongin-si, Gyeonggi-do, 17104, Korea; 
	    mihwajin.sf@gmail.com, jeongeun.lee@khu.ac.kr\\
       $^2$ Korea Astronomy and Space Science Institute, 776 Daedeokdae-ro, Yuseong-gu, Daejeon 34055, Korea; ktkim@kasi.re.kr\\
       $^3$ Department of Astronomy, University of Texas at Austin, 1 University Station, C1400, Austin, Texas 78712-0259; nje@astro.as.utexas.edu
}

\begin{abstract}

We analyzed both HCN $J$=1--0 and HNC $J$=1--0 line profiles to study the inflow motions in different evolutionary stages of massive star formation: 54 infrared dark clouds (IRDCs), 69 high-mass protostellar object (HMPOs), and 54 ultra-compact HII regions (UCHIIs). The inflow asymmetry in HCN spectra seems to be prevalent throughout all the three evolutionary phases, with IRDCs showing the largest excess in blue profile. In the case of HNC spectra, the prevalence of blue sources does not appear, excepting for IRDCs. We suggest that this line is not appropriate to trace inflow motion in evolved stages of massive star formation because the abundance of HNC decreases at high temperatures. This result spotlights the importance of considering chemistry in the dynamics study of massive star--forming regions. The fact that the IRDCs show the highest blue excess in both transitions indicates that the most active inflow occurs in the early phase of star formation, i.e., the IRDC phase rather than in the later phases. However, mass is still inflowing onto some UCHIIs. We also found that the absorption dips of the HNC spectra in 6 out of 7 blue sources are red--shifted relative to their systemic velocities. These red-shifted absorption dips may indicate global collapse candidates, although mapping observations with better resolution are needed to examine this feature in more detail.

\end{abstract}


\section{Introduction}

Massive stars are considered to be decisive players in the physical and chemical evolution of galaxies, injecting energetic feedbacks into their surroundings. In recent years, there have been many studies trying to examine the formation mechanism of high--mass stars, suggesting an evolutionary sequence of massive star formation as follows. First, the formation of massive stars begins in an infrared dark cloud~(IRDC) identified as dark extinction features against the bright Galactic mid-infrared background~\citep{egan98, simon06a}. Their cold ($\la$25~K) and dense ($\ga 10^5~{\rm cm}^{-3}$) properties and strong (sub)millimeter emissions (e.g., Rathborne et al. 2006) suggest the regions as ideal brithplaces of massive stars. The central condensation then begins heating its environment, evolving to become a high--mass protostellar object~(HMPO). They are luminous infrared point--like sources~($L_{\rm bol} \geq 10^3~L_\odot$) without associated radio continuum emission~\citep{molinari96, molinari00, sridharan02, beuther02}. This protostar continues to gain mass and evolves to produce UV photons, ionizing the gas, becoming an ultracompact HII region~(UCHII). UCHIIs are very small (D $\lesssim$ 0.1 pc), dense ($n_{\rm e} \gtrsim10^4~{\rm cm}^{-3}$), and bright ($EM \gtrsim10^7~{\rm pc~cm}^{-6}$) ionized regions ~\citep{woodchurchwell89, kurtz94, ktkim01}. The objects are considered to represent the childhood of HII regions.

Nevertheless, the formation mechanism for the massive stars is still in debate. There are two competing theories describing the massive star formation: turbulent core accretion and competitive accretion~\citep{mckeeandostriker07, zinneckerandyorke07, bodenheimer11}. In the turbulent core accretion model, high--mass cloud cores form from a much larger molecular cloud clump, which is supported by quasi--virialized turbulent flows. The material that ends up as stars can be essentially determined by the process of fragmentation of the cloud clump because the cores are almost non-interacting and the remainder of the clump seldom affects the inflowing process~\citep{mckeeandtan03}. This scenario well explains an initial mass function similar to a core mass function, which is consistent with observations~\citep{motte98, beutherandschilke04, krumholzandtan07}.

In the competitive accretion model, star formation is regulated by the global collapse of a much larger cloud, initially containing gas of several thousand $M_\sun$. The material that ends up as stars is gathered during the star-formation process from various parts of the parent cloud. The cores compete for the remaining gas and there are strong interactions among them. This scenario predicts that massive stars form at the cluster center where more massive inflow can occur than in the outer regions~\citep{bonnell01, krumholzandbonnell09}.

 Regardless of which mechanism works, gravitational inflow is a key process to initiate star formation and to control the evolution of densities in the protostellar envelope. Therefore, characterizing this inflow process is important for a better understanding of high-mass star formation. One observational signature of inflow motion is a `blue profile', a general prediction for a cloud collapsing model~\citep[e.g.][]{shu77}. This blue profile is an asymmetric line feature that appears in an optically thick line profile with a self-absorption dip and a blue peak stronger than a red peak. The emission of an optically thin line peaks near the absorption dip of the optically thick line.

There have been many attempts to examine the inflow signature in massive star--forming regions in recent years. After \citet{wuandevans03} found statistically significant blue excess (the number of blue profile minus that of red profile in units of the total number of sample) in the blue profile in the HCN $J$=3--2 line toward early phase of HII regions where star-forming activity still appears, the number of studies has been increasing. For example, \citet{reiter11a} carried out HCO$^{+}$ $J$=3--2 line observations toward similar regions of \citet{wuandevans03}. They noted that every source with the blue asymmetry in the HCO$^{+}$ $J$=3--2 line also has the blue profile in the HCN $J$=3--2 line, and confirmed that the HCN $J$=3--2 line is a better inflow tracer. \citet{rygl13} performed the HCO$^{+}$ $J$=1--0, 4--3, and CO $J$=3--2 line observations toward a sample of clumps in clouds with high extinctions, and they concluded that among the three transitions, the HCO$^{+}$ $J$=1--0 line is the most sensitive to detect inflowing motions. In addition, \citet{fuller05} reported significant excess of blue profiles toward HMPOs in the HCO$^{+}$ $J$=1--0, 3--2, and 4-3 transitions and H$_2$CO $2_{12}$--$1_{11}$ line.

In addition, there have been several previous studies dealing with evolutionary tendency of inflow motion. For example, \citet{wu07} showed dramatic increase of blue excess in the HCO$^{+}$ $J$=1--0 line with evolution from HMPOs to UCHIIs. Other studies have found different results. \citet{purcell06} revealed equal numbers of red and blue profiles of the HCO$^{+}$ $J$=1--0 line toward HMPOs and UCHIIs, and found blue excess only in IRDCs, indicating active inflow motion occurring in the early phase of star formation. Recently, an extensive inflow survey toward 405 compact sources classified into prestellar, protostellar and UCHII regions was performed in the HCO$^{+}$ $J$=1--0 and HNC $J$=1--0 lines~\citep{he15}. They suggested that the HCO$^{+}$ $J$=1--0 line is better to trace inward motion and found that the blue excess declines with evolutionary stage. With a higher transition of HCO$^{+}$, the opposite tendency appears. In the HCO$^{+}$ $J$=4--3 study by \citet{klaassen12}, 12 out of 22 UCHIIs show the blue asymmetric line profile while only 3 blue sources are detected among 12 HMPOs. They attribute this lower occurrence of blue profiles in HMPOs to the beam dilution effect.

Except for the study of \citet{wu07}, all of these results regarding HCO$^{+}$ $J$=1--0 line can be summarized as follows; (1) the HCO$^{+}$ $J$=1--0 transition is likely to be the most sensitive inflow tracer. (2) The blue excess measured with this line intensity decreases with evolution of the massive star--forming regions, suggesting that the younger the sources are, the easier to detect inflow with this tracer. (3) This observing trend, however, can appear differently with the higher transition lines.

However, many of those studies mainly used the HCO$^{+}$ transitions as an inflow tracer or dealt with limited phases of massive star formation. However, different line transitions at different molecular species must be tested because inflow could be associated with various excitation conditions. In this study, we search for inflow candidates toward various evolutionary stages related to massive star formation~(IRDCs, HMPOs, and UCHIIs) using the HCN and HNC $J$=1--0 lines. This paper is organized as follows. The details about source selection and observations are provided in \S 2. Analyses for asymmetric profiles are presented in \S 3. The discussion for inflow candidates is provided in \S 4. The main results are summarized in \S 5.

\section{Observation}

\subsection{Target Selection} 

After \citet{rathborne06} identified 190 compact cores in the 1.2~mm continuum images of the 38 darkest IRDCs, \citet{chambers09} classified them as `quiescent' prestellar cores~(qIRDCc) and `active' protostellar cores~(aIRDCc). aIRDCc show both 4.5 and 24 $\micron$ infrared emission which is the signature of star--forming activities, while qIRDCc contain neither emission. We adopted 19 qIRDCc and 35 aIRDCc from the catalog of \citet{chambers09} as our IRDC targets. We selected 69 HMPOs from the catalogs of \citet{sridharan02} and \citet{molinari96} and 54 UCHIIs from the catalogs of \citet{woodchurchwell89} and \citet{kurtz94}. The details about selection criteria are provided in \citet{jin15}. Consequently, our sample consists of 54 IRDCs (19 qIRDCc and 35 aIRDCc), 69 HMPOs, and 54 UCHIIs.

\subsection{Observation} 

The $J$=1--0 transitions of HCN and HNC and their isotopic lines~(Table~\ref{lineproperties}) were observed in 2012--2013 using the Korean VLBI Network (KVN) 21m telescope at the Yonsei and Ulsan stations~\citep{ktkim11, lee11}. The main--beam efficiencies are 0.43 and 0.37 for the KVN Yonsei and Ulsan telescopes, respectively, and the beam sizes of both telescopes are 32 $\arcsec$. All the lines were observed with the position switching mode and their intensities were calibrated on the ${T_\textrm{A}}^{*}$ scale by the standard chopper wheel method. The focus and pointing were adjusted by observing strong SiO maser sources every one to two hours. The system temperature ranged from 170~K to 280~K. The rest frequencies, dipole moments, and relative weights of the hyperfine components of the observed lines are summarized in Table~\ref{lineproperties}. All spectra were reduced using CLASS in the GILDAS software package, and the reduced line spectra have velocity resolution of 0.21~\kms .

\section{ Analysis \& Results }

To select sources for analysis of inflow signatures, first, we used a 3$\sigma$--detection criterion for each line. After that, some sources are excluded by eye if they are suspected to have multiple velocity components in a line of sight. Specifically, eight IRDC cores are excluded in the analysis of HCN $J$=1--0. The hyperfine components are strongly self-absorbed and blended each other, making it hard for the lines to be exploited in the inflow analysis. In addition, there are additional emission components that cannot be solely explained by the combination of self-absorption and line blending effects in some sources. Finally, 12 IRDCs, 26 HMPOs and 23 UCHIIs are selected for the HCN and \hcn\ line analysis while 25 IRDCs, 28 HMPOs and 23 UCHIIs are selected for the analysis of the HNC and \hnc\ lines. The information of the sources selected for analysis is listed in Table~\ref{sourceinform}.

According to \citet{wuandevans03}, the inflow proceeds at a relatively low velocity so that its observational signature can be easily masked by other mechanism or a beam dilution effect. However, all our samples that are listed in the SCUBA legacy catalogue have larger effective radii than half of our beam size in the 850$\micron$ continuum~\citep{james08, jin15}, indicating that the emission is not likely beam--diluted. The HCN and HNC line emissions are known to be well correlated with the dust emission~\citep{wu10, reiter11b}.

\begin{figure}
\centering
\includegraphics[scale=0.7]{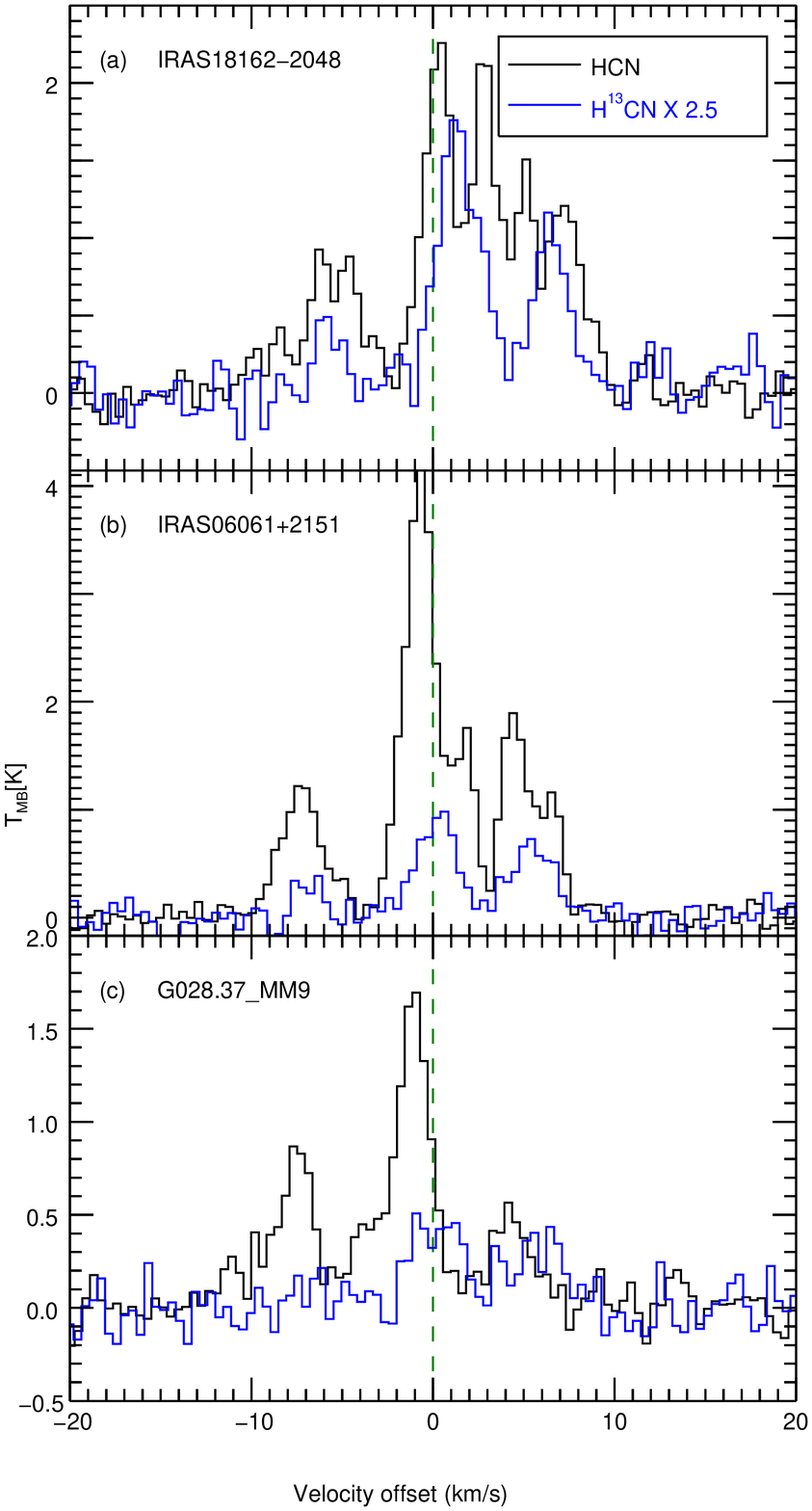}
\caption{Various features of blue profiles in the HCN $J$=1--0 line. Both lines are plotted with the velocity relative to the optically thin line's central velocity~(green dashed lines). }
\label{blueprofileexamplehcn}
\end{figure}

\begin{figure}
\centering
\includegraphics[scale=0.7]{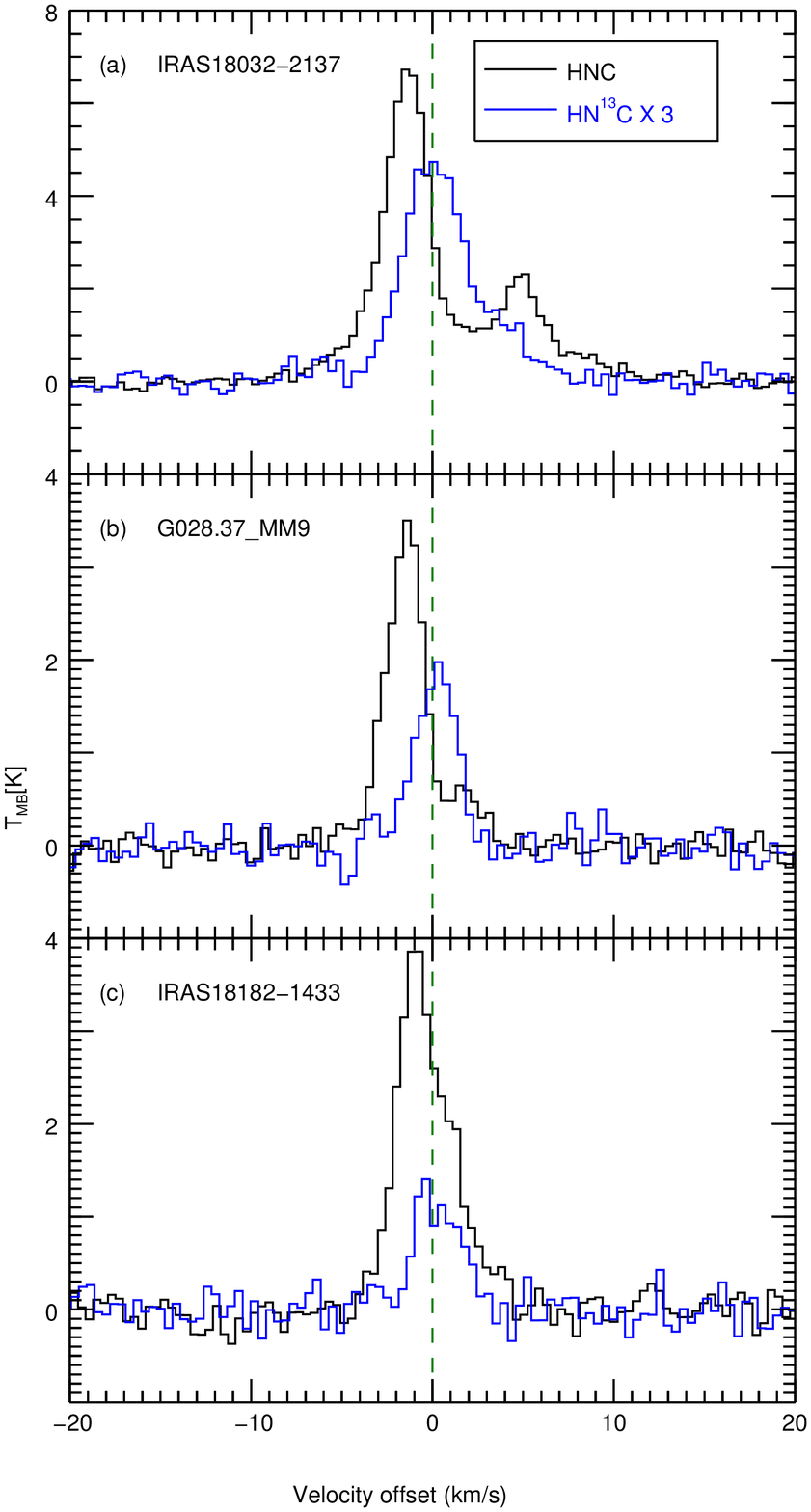}
\caption{Various features of blue profiles in the HNC $J$=1--0 line. Both lines are plotted with the velocity relative to the optically thin line's central velocity~(green dashed lines). }
\label{blueprofileexamplehnc}
\end{figure}

The general signature of inflow is the so called `blue profile'. This is an asymmetric line feature with a self-absorption dip where the blue peak is stronger than the red peak, while an optically thin line must peak near the dip of the optically thick line. In this case, the ratio of the blue peak to the red peak~($T(B)/T(R)$) can be one measure for the line asymmetry. However, depending on the opacity of the line, the blue profiles can show other features, for example, a single blue peak with red shoulder or a blue-skewed single peak. Figure~\ref{blueprofileexamplehcn}--\ref{blueprofileexamplehnc} show the various features of blue profiles in the HCN and HNC $J$=1--0 lines, respectively, ranging from a clearly self-absorbed blue profile to a blue-skewed profile.

In low--mass star forming regions, \citet{mardones97} have suggested $ \delta v $ as an alternative measure of the blue profile for these blue skewed lines, which is defined as a difference between the line central velocity of an optically thick line~($ v_\textrm{thick} $) and that of an optically thin line~($ v_\textrm{thin} $), in units of the line width of the optically thin line~($ \Delta v_\textrm{thin} $) 
\begin{equation}
\delta v = \frac{v_\textrm{thick}-v_\textrm{thin}}{\Delta v_{\textrm{thin}}}.
\end{equation}
A line can be identified as blue/red profile if the difference between $ v_\textrm{thick} $ and $ v_\textrm{thin} $ is greater than a quarter of $ \Delta v_\textrm{thin} $. That is, a blue profile would have \deltav\ $<$ -0.25 while a red profile would have \deltav\ $>$ 0.25~\citep{mardones97}. However, it is important to note that adopting the same boundaries on \deltav\ for these high--mass objects actually demands a larger velocity shift than for the low--mass sources because the molecular lines towards these high--mass samples are significantly broader~\citep{fuller05}.

We measured the line asymmetries of all the detected sources using the \deltav\ analysis under the assumption that both \hcn\ and \hnc\ lines are optically thin. The optical depth for each line was obtained by adopting the values of \citet{jin15} or by following the same analysis described therein. The resulting mean value for each line was less than 0.12 in all the evolutionary stages. For the HCN and \hcn\ $J$=1--0 spectra consisting of three apparent hyperfine lines, the strongest hyperfine component ($F$=2--1) is adopted as a standard for \deltav\ calculation. All the line central velocities~($ v_\textrm{thick} $, $ v_\textrm{thin} $) and line widths~($ \Delta v_\textrm{thin} $) are determined from multiple--Gaussian fitting. The HNC $J$=1--0 line also has a hyperfine structure, but the splitting is too small~($\thicksim$ 0.7 \kms; \citet{vandertak09}) to perform the multiple Gaussian fitting. As a result, the values of \deltav\ were calculated in the same manner, but all the line parameters are measured with a single--Gaussian fitting.

\begin{sidewaysfigure}
\includegraphics[scale=0.6]{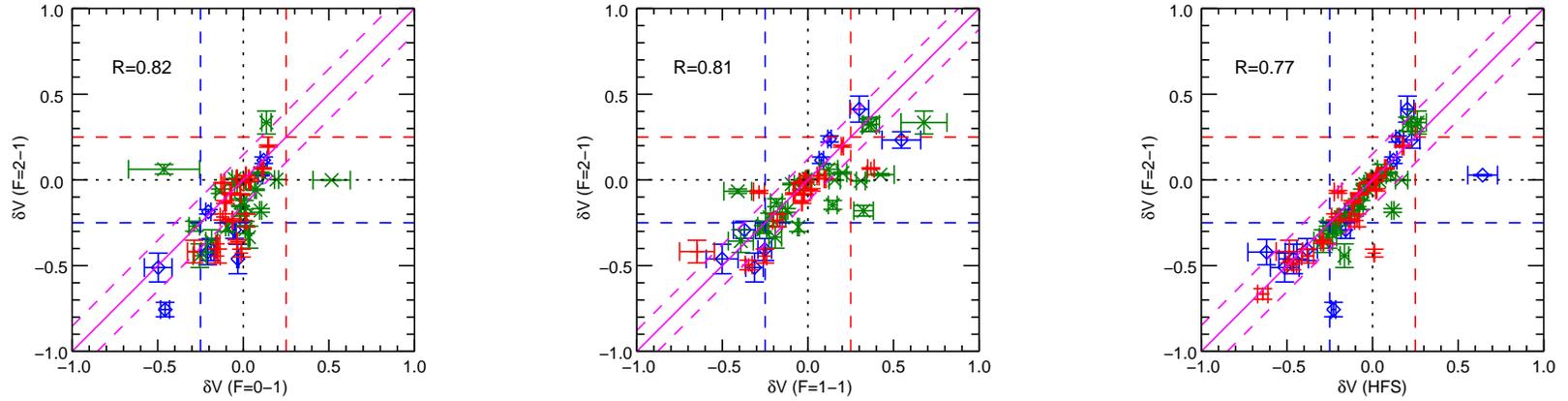}
\caption{Comparison the \deltav\ values derived from the main hyperfine component~($F$=2--1) of the HCN $J$=1--0 line with the values derived from the other hyperfine satellites~(left and medium panel) and from the all three hyperfine components at once~(right panel). IRDC cores, HMPOs, and UCHIIs are indicated by diamonds, times, and crosses, respectively. The solid line is the line of perfect correlation and the dotted lines indicate 1--$\sigma$ from the line. The Pearson correlation coefficients~(R) are given in the upper left corner of each panel, and $p$--values for all correlations are extremely small. The number of samples in each box is 50, 49, and 61, respectively, from left to right. One HMPO~(IRAS23140+6121) whose value is (\deltav\ ($F$=0--1)=-1.606, \deltav\ ($F$=2--1)=1.710) is not presented in the first panel.}
\label{HFSprofilecomparison}
\end{sidewaysfigure}

\begin{figure}
\centering
\includegraphics[scale=0.55]{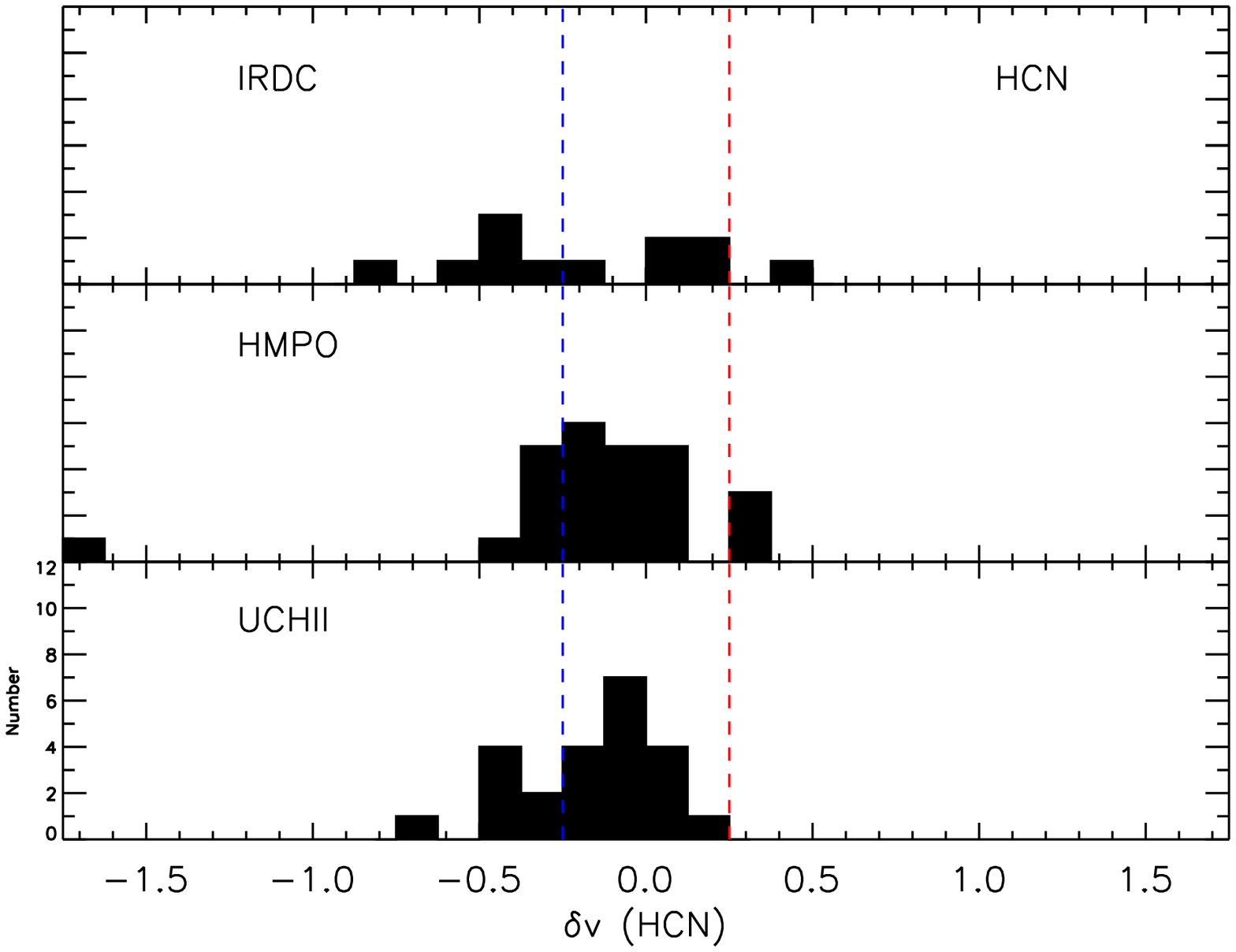}
\caption{Distribution of \deltav\ values calculated from the HCN $J$=1--0 line. The sources located outside the blue/red dashed lines can be considered as blue/red profiles, whereas the sources inside the blue and red dashed lines are neutral profiles. The top, middle, and bottom panels represent results from 12 IRDC cores, 26 HMPOs, and 23 UCHIIs, respectively.}
\label{HCNdelvdistribution}
\end{figure}

\begin{figure}
\centering
\includegraphics[scale=0.55]{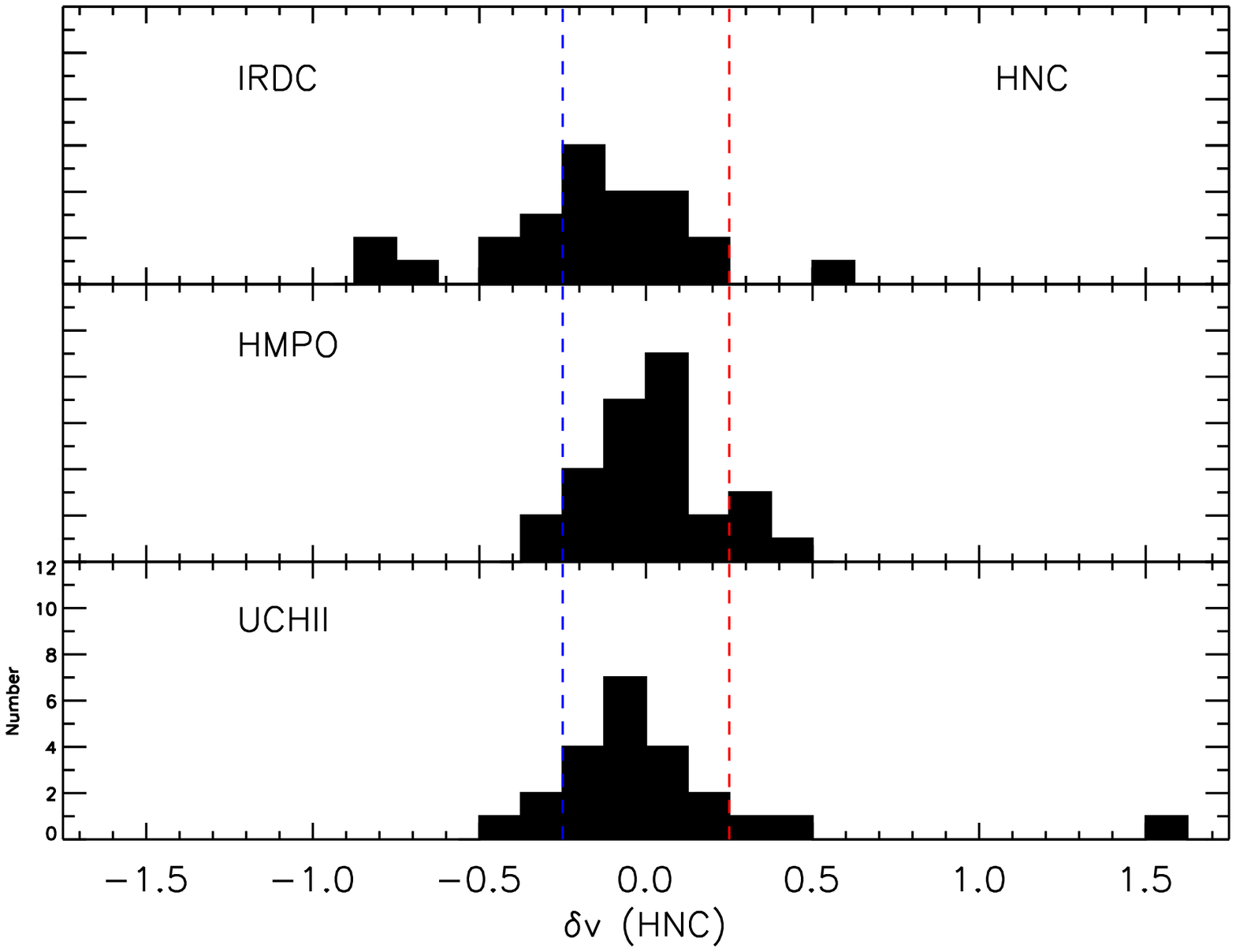}
\caption{Distribution of \deltav\ values calculated from the HNC $J$=1--0 line. The sources located outside the blue/red dashed lines can be considered as blue/red profiles, whereas the sources inside the blue and red dashed lines are neutral profiles. The top, middle, and bottom panels represent results from 25 IRDC cores, 28 HMPOs, and 23 UCHIIs, respectively.}
\label{HNCdelvdistribution}
\end{figure}

As mentioned above, the strongest hyperfine component~($F$=2--1) is mainly used for the HCN $J$=1--0 transition in the \deltav\ analysis because this component is detected toward all sources, unlike other weak components, so that we can maximize the sample number for our analysis. Prior to adopting the $F$=2--1 hyperfine component as standard, we derived the correlations among \deltav\ values not only measured from the Gaussian fitting for each HCN hyperfine component but also measured by fitting the whole hyperfine structure simultaneously. As presented in Figure~\ref{HFSprofilecomparison}, they show tight correlations in the confidence level above 99~\%. 
Therefore, $F$=2--1 can be representative for all hyperfine components in our analysis. The observed line parameters and derived \deltav\ are listed in Tables~\ref{hcnlinepara} and \ref{hnclinepara}. 
Figures~\ref{HCNdelvdistribution} and \ref{HNCdelvdistribution} show the distribution of \deltav\ derived from HCN and HNC lines, respectively. The sources located in the left side of the blue dashed line have blue profiles, while those on the right side of the red dashed line have red profiles.

An asymmetric profile may also be induced by other mechanism~(e.g., rotation and outflow). If this is the case, a large sample with a random distribution of angles between the axis and the line of sight will not produce an excess of one type of profile~\citep{wuandevans03}.
So the concept of the ``blue excess" was introduced by \citet{mardones97} to quantify the statistics of the line asymmetry in a survey:
\begin{equation}
E=\frac{N_\textrm{blue}-N_\textrm{red}}{N_\textrm{total}}
\end{equation}
where $N_\textrm{blue}$ and $N_\textrm{red}$ are the numbers of blue and red profiles in the total samples~($N_\textrm{total}$). These statistical results are summarized in Table~\ref{statresult}.

\section{ Discussion }

Values of the blue excess derived from the HCN $J$=1--0 line for IRDCs, HMPOs, and UCHIIs~(0.42, 0.15, and 0.30, respectively) are larger than those derived from the HNC $J$=1--0 line~(0.28, -0.07, and 0.00, respectively). In the HCN spectra, a prevalence of blue profiles relative to red profiles is found in every evolutionary stage, with the IRDCs showing the largest blue excess~(Figure~\ref{HCNdelvdistribution}). In contrast, the distribution of the \deltav\ derived from the HNC line is relatively centered on neutral profiles, excepting for IRDCs~(Figure~\ref{HNCdelvdistribution}).

We performed a binomial test and calculated the probability $P$ that the one type of excess is induced by chance. The binomial distribution is described as follows
\begin{equation}
P = \binom {n}{k} p^{k} (1-p)^{(n-k)}\,,
\end{equation}
where $n$ is the total number of trials, $k$ is the number of successes, and $p$ is the success probability. In this case, $n$ is the total number of sources, $k$ is the number of the blue sources, and the success probability $p$=0.5 if the distribution shows no bias toward red or blue. Then the possibility that the number of blue sources is equal to or higher than the observed number by chance can be calculated by adding all possibilities $P(n, k, p) + P(n,k+1,p) + ...$ until $k=n$. A small value of $P$ indicates that it is unlikely for a blue excess to arise by chance~\citep{rygl13}. All resulting values of $E$ and $P$ are listed in Table~\ref{statresult}. The $E$ of the HCN line is statistically significant with a sufficiently low probability $P$ throughout all evolutionary phases. In the case of the HNC line, in contrast, such a significant value of $E$ appears only in the IRDC phases. The fact that the IRDCs show the highest blue excess in both inflow tracers indicates that the most active inflow occurs in the early phase of massive star formation, even though the characteristics of blue profile largely depend on the suitable combination of optical depth and critical density. It should be noted that the small sample size of HCN sources in IRDCs would bring about statistical instability in calculating blue excess $E$. Nevertheless, the probability $P$ as low as 6 $\%$ indicates that the prevalence of blue profile is not likely to occur by chance. In addition, the HNC line also shows significant excess to blue in the IRDCs.

\subsection{The astrochemical effect on inflow tracer}

\begin{figure}
\centering
\includegraphics[width=\textwidth]{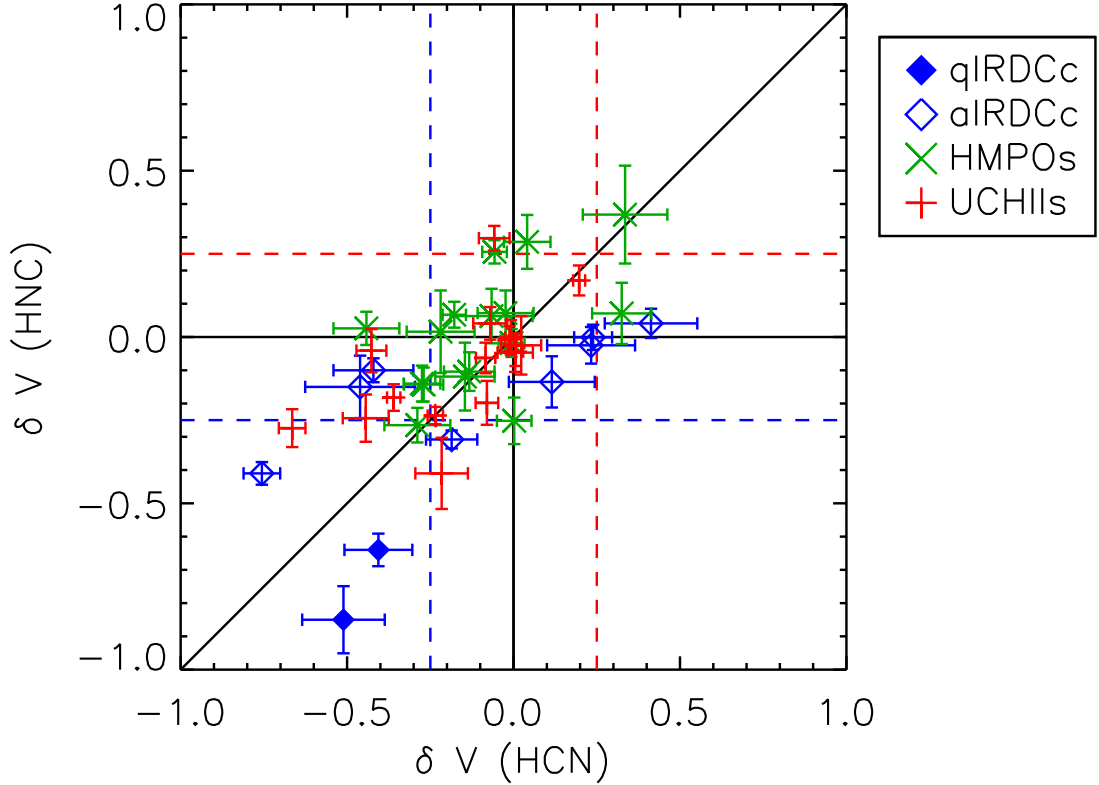}
\caption{Comparison of \deltav\ values between HNC and HCN. Only the sources where we can analyze both inflow tracers are plotted here. IRDC cores, HMPOs, and UCHIIs are indicated by diamonds, times, and crosses, respectively. Here IRDC cores are divided into quiescent ones (qIRDCc, filled diamonds) and active ones (aIRDCc, open diamonds), depending on star-forming activity. The sources located outside the blue/red dashed lines can be considered as blue/red profiles. One UCHII source whose value is (-0.137, 1.524) is not presented here. }
\label{HNCdelvVSHCNdelv}
\end{figure}

These results suggest that the HCN $J$=1--0 line is a better inflow tracer than the HNC $J$=1--0 line in massive star--forming regions. The \deltav\ values of the sources detected in both inflow tracers are plotted in Figure~\ref{HNCdelvVSHCNdelv}. The sources located outside the blue/red dashed lines are considered as the blue/red profiles, while the sources located inside those lines are regarded as neutral profiles. Many sources blue in HCN are neutral in HNC, but not vice versa, indicating the HNC is less appropriate to trace inflow motion. We attribute this to an astrochemical effect that reduces the abundance, hence, the optical depth of HNC.

\citet{jin15} have found that HCN/HNC abundance ratio increases while the optical depth of \hnc\ decreases as sources evolve from IRDC to UCHIIs, even though both HCN and HNC are mainly formed in equal measure by dissociative recombination~\citep{mendes12}. One suggested reason for this phenomenon is a neutral--neutral reaction where HNC is selectively consumed at high temperatures~($T_\textrm{K} \ge$ 24~K;~\citet{hirota98}). \citet{hirota98} showed that the HCN abundances in the high kinetic temperature regions (OMC-1 cores) are comparable to those in the dark cloud cores whereas the HNC abundances decrease as the temperature rises. By this astrochemical effect, the opacity of the HNC line would decrease as an object evolves so that the line cannot trace inflow motion well; an inflow profile appears in lines sufficiently opaque~\citep{myers96}. In Figure~\ref{HNCdelvVSHCNdelv}, a significant number of HMPO and UCHII sources are bluer in HCN than HNC. On the contrary, the HNC line is rather bluer in the IRDCs, and this opposite tendency is more obvious in qIRDCc than aIRDCc, supporting our scenario again. The qIRDCc is considered to be in the earlier phase~\citep{chambers09} and show the smaller value of the HCN/HNC abundance ratio than aIRDCc~\citep{jin15}. This result spotlights the importance of regarding chemistry when studying dynamics of star--forming regions.

\subsection{Comparison with previous studies}

There have been many attempts to examine the inflow signature as mentioned in \S 1. Some of those studies have suggested the HCO$^{+}$ $J$=1--0 line as the best inflow tracer in massive star--forming regions. We compare our results not only with the previous inflow surveys using the HCO$^{+}$ $J$=1--0 line but also with the study using a higher transition line of HCN.

In the HCO$^{+}$ $J$=1--0 line, IRDCs seem to undergo the most detectable active inflow process ~\citep{purcell06, rygl13, he15} whereas sources in more evolved phases such as HMPOs and UCHIIs show less inflow as indicated by smaller values of the blue excess. For example, \citet{purcell06} reported the blue excess as low as 0.02 toward these evolved samples, and \citet{he15} found the decreasing tendency of the blue excess with evolution from IRDCs to UCHIIs. The blue excess that we observed with the HCN $J$=1--0 line are comparable with their values for each evolutionary stage with the largest value in IRDCs. This indicates that the HCN $J$=1--0 line is as sensitive as the HCO$^{+}$ $J$=1--0 line in high--mass star forming regions.

For a higher transition of HCN, \citet{wuandevans03} surveyed inflow motion using the HCN $J$=3--2 line. They reported the blue excess of 0.21 in the sources consisting of the 28 HII regions where the star-forming activity still appears. This value is much smaller than the blue excess for our UCHIIs sample~(0.30 for 23 UCHIIs), showing that the 1-0 transition line of HCN traces the inflow motion better than the higher transition line. This result is consistent with the result of \citet{fuller05}; the lower transition lines of HCO$^{+}$ show the more inflow signature. They observed the HCO$^{+}$ $J$=1--0, 3--2, and 4--3 transitions toward HMPOs and found the highest blue excess in the $J$=1-0 line. This result may be related to the gas motion~(i.e., velocity profile) that the high energy level transitions trace; the higher transition lines emit from the hotter and denser central region. According to a model of collapsing cloud in massive star--forming regions, the velocity gradient at the central region is too large to make the self--absorption feature in high energy transitions~\citep{smith13}.

\subsection{Global collapse?}

\textit{An interesting feature in double--peaked HNC spectra is that the absorption dips in 6 out of 7 blue sources are red--shifted relative to the systemic velocities.} For the sources whose lines are strongly self--absorbed, the asymmetries of the spectra are determined using the $T(B)/T(R)$ parameter. The fluxes of the two peaks are measured by the double-Gaussian fitting, and if the differences between the peaks are larger than the 3$\sigma$ noise level, we classify them into blue/red profiles otherwise neutral profile. Some sources suspected to have a wing-like structure, however, couldn't be fitted by the double-Gaussians directly even though an obvious self-absorption feature appears. Those lines are fitted after masking the wing-like structures. After that, we compare the velocity of the absorption dip~($v_\textrm{dip}$) with the systemic velocity determined from the \hnc\ line. The $v_\textrm{dip}$ is identified as a velocity at the lowest flux in the absorption dip by cursor. If the velocity deviation of the absorption dip exceeds three times the measurement error of the systemic velocity, the line is considered to have a shifted dip. 
The asymmetry parameters ($T(B)/T(R)$), the velocities of the absorption dips~($v_\textrm{dip}$) of the HNC $J$=1--0 line and central velocities of the HN$^{13}$C $J$=1--0 line ($v_\textrm{thin}$) are listed in Table~\ref{hnclinepara}. We also tried to perform the same analysis in the HCN spectra. However, the double--Gaussian fitting was not reliable because of the combination of the line blending effects among the hyperfine components and the self-absorption.

\begin{figure}
\centering
\includegraphics[width=\textwidth]{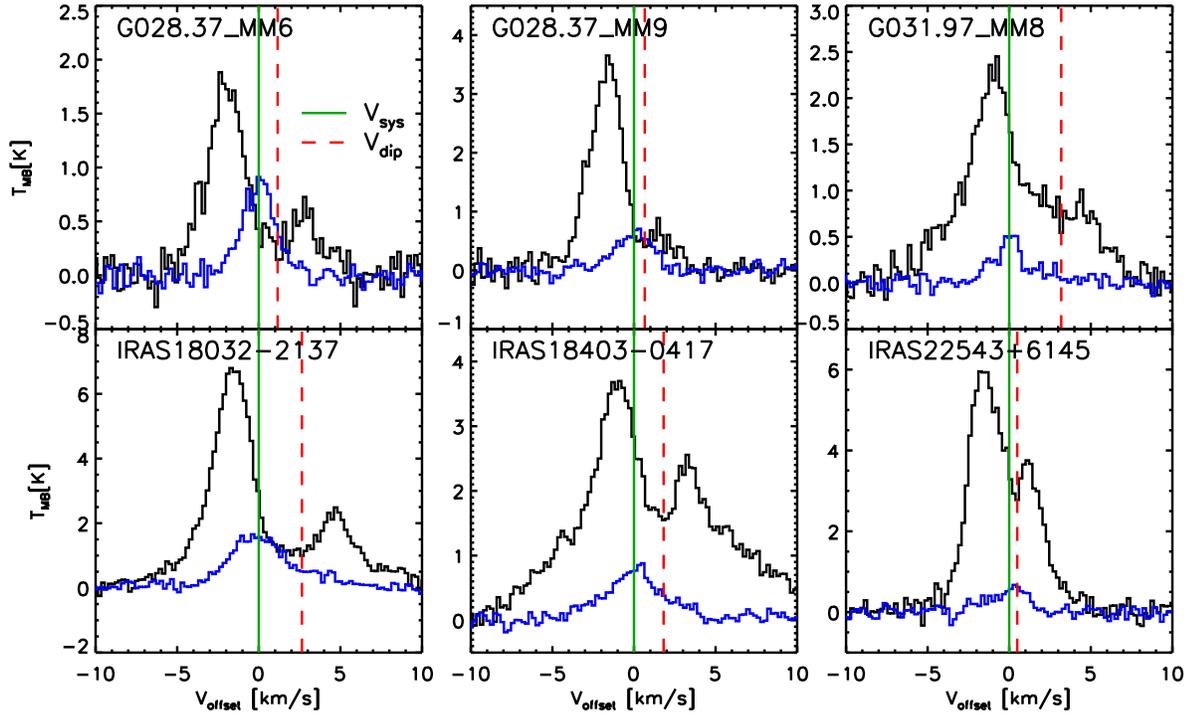}
\caption{The blue profile with red-shifted absorption dip in the HNC $J$=1--0 line. Both HNC~(black) and \hnc\ (blue) lines are plotted with the velocity relative to the systemic velocity~(green solid lines). Red dashed line represents the dip position.}
\label{overallcollapseeg}
\end{figure}

According to the above analysis, 6 out of 7 blue sources have absorption dips red--shifted relative to their systemic velocity~(Figure~\ref{overallcollapseeg}). If considering turbulent core accretion model, the star formation occurs in quasi-equilibrium molecular cloud where inflow occurs in localized regions. This would make the absorption dip at the source velocity. However, if the cloud clumps form in global collapse as described in the competitive accretion model, even the outer larger region takes part in the inflowing process, making the absorption dip red--shifted. Therefore, these red-shifted absorption dips detected in our sources may indicate global collapse candidates. Actually, \citet{smith13} calculated the line profiles of HCO$^{+}$ in a core following the competitive accretion formalism and frequently found non--central self--absorption dip. However, depending on optical depth, red-shifted absorption dip can be also induced by the absorption in the inner collapsing regions. Mapping observations with better resolution are needed to rule out this possibility.

\section{Summary}

To understand the gravitational inflow taking place in high--mass star formation, we surveyed 54 IRDC cores, 69 HMPOs, and 54 UCHIIs in the HCN $J$=1--0 and HNC $J$=1--0 lines. 

(1) We found a statistically significant blue excess of the HCN line for every evolutionary phase~(0.42, 0.15, and 0.30 for IRDCs, HMPOs, and UCHIIs, respectively). These are comparable to the values derived using other inflow tracers, including HCO$^{+}$ $J$=1--0, known to be one of the best inflow tracer. This indicates the HCN line is a good tracer of gravitational inflow. 

(2) With the HNC line, the blue profile appears significant only in IRDCs. We concluded that this line is not appropriate to trace inflow motion in evolved stages of massive star formation because of the HNC abundance~(and thus, its optical depth) decreases at high temperatures. This result spotlights the importance of considering chemistry in studying dynamics of massive star--forming regions.

(3) The fact that IRDCs show the highest blue excess in both inflow tracers indicates that the IRDC phase is undergoing the most active inflow process. This result is consistent with a general prediction of inflow process where the younger sources are expected to be more actively inflowing onto the central source. However, the UCHIIs is also likely inflowing matters yet. 

(4) We found that the absorption dips of the HNC $J$=1--0 spectra are red--shifted relative to the systemic velocities in 6 out of 7 blue sources. These red-shifted absorption dips suggest that the clumps are in global collapse. Mapping observations with better angular resolutions are needed to examine this feature in more detail. 

\section*{Acknowledgements}

We are grateful to all staff members in KVN who helped to operate the telescope. The KVN is a facility operated by the Korea Astronomy and Space Science Institute. This work was supported by the Basic Science Research Program through the National Research Foundation of Korea (NRF) (grant No. NRF-2015R1A2A2A01004769) and the Korea Astronomy and Space Science Institute under the R $\&$D program (Project No. 2015-1-320-18) supervised by the Ministry of Science, ICT and Future Planning.

\clearpage

\clearpage

\clearpage

\clearpage

\clearpage

\clearpage

\clearpage

\begin{deluxetable}{lclll}
\tablewidth{0pt}
\tabletypesize{\footnotesize}
\tablecolumns{5}
\tablecaption{Observed Lines}
\scriptsize
\tablehead{
\colhead{Molecule} 		& \colhead{Transition}	 	& \colhead{$\nu$ [MHz]}	 & \colhead{$\mu$ [D]}		& \colhead{$S_\textrm{ul}^\textrm{a}$}
}
\startdata
HCN & $J$=1-0, F=2--1 & 88630.42 & 2.99$^\textrm{b}$ & 3\\
 &$J$=1--0, F=1--1 & 88631.85 &  & 5\\
 & $J$=1--0, F=0--1 & 88633.94 &   & 1\\
\hcn\ & $J$=1-0, F=2--1 & 86338.77 & 2.99$^\textrm{b}$ & 3\\
 &$J$=1--0, F=1--1 & 86340.18 &  & 5\\
 & $J$=1--0, F=0--1 & 86342.27 &   & 1\\
HNC & $J$=1--0 & 90663.57 & 3.05$^\textrm{c}$  & \\ 
\hnc\ & $J$=1--0 & 87090.85 & 3.05$^\textrm{c}$  & \\
\enddata
\label{lineproperties}
\tablecomments{$^\textrm{a}$ Relative weights of the hyperfine components. $^\textrm{b}$ \citet{bhattacharyaandgordy60}. $^\textrm{c}$ \citet{blackman76}.}
\end{deluxetable}

\begin{deluxetable}{clllllllc}
\tablewidth{0pt}
\tabletypesize{\footnotesize}
\tablecolumns{9}
\tablecaption{Source information}
\scriptsize
\tablehead{
\colhead{Classification}	& \colhead{Source name} 		& \colhead{R.A.}	 	& \colhead{Decl.}	& \colhead{$l$}	 	& \colhead{$b$}	  & \colhead{$D_\textrm{far}$}	& \colhead{$D_\textrm{near}$}	& \colhead{inflow tracer}\\
	&	& \colhead{J(2000.0)}  & \colhead{J(2000.0)}  & \colhead{($\arcdeg$)} & \colhead{($\arcdeg$)} & \colhead{(kpc)} & \colhead{(kpc)} &
}
\startdata
qIRDCc	&                G028.37$\_$MM9    & 18:42:46.7      &           -04:04:08     & 28.32    &  0.07    &  5.0    &	\nodata	& HCN, HNC\\[-1ex]
		&                G031.97$\_$MM9    & 18:49:31.6      &           -00:46:30     & 32.02    &  0.07    &  6.9    &	\nodata	& HNC\\[-1ex]
 		&                G035.39$\_$MM5    & 18:57:08.8      &           +02:08:09    & 35.48    & -0.30    &  2.9    &	\nodata	& HCN, HNC\\[-1ex]
aIRDCc	&                G015.31$\_$MM3	& 18:18:45.3	&		-15:41:58	  & 15.28	& -0.09	& 3.2	   &	\nodata	& HNC\\[-1ex]
		&                G022.35$\_$MM1    & 18:30:24.4      &           -09:10:34     &22.38    &  0.45    &  4.3    &	\nodata	&  HCN\\[-1ex]
		&                G023.60$\_$MM1    & 18:34:11.6      &           -08:19:06     &23.57    &  0.01    &  3.9    &	\nodata	&  HNC \\[-1ex]
		&                G023.60$\_$MM6    & 18:34:18.2	  &		-08:18:52     &23.59    &  -0.01    &  3.9    &	\nodata	&  HNC\\[-1ex]
		&                G024.60$\_$MM1    & 18:35:40.2      &           -07:18:37     &24.63    &  0.15    &  3.7    &	\nodata	& HNC \\[-1ex]
		&                G028.37$\_$MM4    & 18:42:50.7      &           -04:03:15     &28.34    &  0.06    &  5.0    &	\nodata	&  HNC\\[-1ex]
		&                G028.37$\_$MM6    & 18:42:49.0      &           -04:02:23     &28.36    &  0.07    &  5.0    &	\nodata	& HNC\\[-1ex]
		&                G030.97$\_$MM1    & 18:48:21.6      &           -01:48:27     &30.97    & -0.14    &  5.1    &	\nodata	& HCN, HNC\\[-1ex]
		&                G031.97$\_$MM1    & 18:49:36.3      &           -00:45:45     &32.04    &  0.06    &  6.9    &	\nodata	&  HCN, HNC\\[-1ex]
		&                G031.97$\_$MM8    & 18:49:29.1	&		-00:48:12     &32.00   &  0.07    &  6.9    &	\nodata	&  HNC\\[-1ex]				
		&                G033.69$\_$MM4    & 18:52:56.4      &           +00:43:08     &33.74    &  0.00    &  7.1    &	\nodata	& HCN, HNC \\[-1ex]
		&                G033.69$\_$MM5    & 18:52:47.8      &           +00:36:47     &33.63    & -0.02    &  7.1    &	\nodata	&  HCN, HNC\\[-1ex]
		&                G034.43$\_$MM1    & 18:53:18.0      &           +01:25:24     &34.41    &  0.24    &  3.7    &	\nodata	& HNC \\[-1ex]
		&                G034.43$\_$MM3    & 18:53:20.4      &           +01:28:23     &34.46    &  0.25    &  3.7    &	\nodata	& HNC \\[-1ex]
		&                G034.43$\_$MM4    & 18:53:19.0      &           +01:24:08     &34.39    &  0.22    &  3.7    &	\nodata	& HCN, HNC \\[-1ex]
		&                G035.39$\_$MM7    & 18:57:08.1      &           +02:10:50     &35.52    & -0.27    &  2.9    &	\nodata	& HCN \\[-1ex]
		&                G035.59$\_$MM3	& 18:57:11.6	&		+02:16:08	   &35.61	& -0.24	& 2.9		&	\nodata	& HNC \\[-1ex]
		&                G038.95$\_$MM3	& 19:04:07.4	&		+05:09:44	   &38.97	&-0.46	& 2.7		&	\nodata	& HNC\\[-1ex]
		&                G038.95$\_$MM4	& 19:04:00.6	&		+05:09:06   & 38.95     &-0.44	& 2.7		&	\nodata	& HNC\\[-1ex]
		&                G048.65$\_$MM1    & 19:21:49.7      &           +13:49:30     &48.67    & -0.30    &  2.5    &	\nodata	& HCN, HNC \\[-1ex]
		&                G048.65$\_$MM2    & 19:21:47.6      &           +13:49:22     &48.66    & -0.30    &  2.5    &	\nodata	& HNC \\[-1ex]
		&                G053.25$\_$MM4    & 19:29:34.5      &           +18:01:39     &53.25    &  0.05    &  1.9    &	\nodata	&  HCN, HNC\\[-1ex]
		&                G053.25$\_$MM6    & 19:29:31.5      &           +17:59:50     &53.22    &  0.05    &  1.9    &	\nodata	& HCN, HNC \\[-1ex]
		&		  G053.31$\_$MM2    & 19:29:42.1     &           +18:03:57     &53.30    &  0.05    &  2.0    &	\nodata	& HNC \\[-1ex]
HMPO	&		 IRAS00117+6412	& 00:14:27.7     &		+64:28:46	 & 118.96   & 1.89 & 1.8	&	\nodata	& HCN\\[-1ex]
		&		 IRAS05358+3543    & 05:39:10.4      &           +35:45:19     &173.48    &  2.43    &  1.8    &  \nodata    & HCN, HNC\\[-1ex]
		&                IRAS05373+2349    &05:40:24.4      &           +23:50:54     &183.72    & -3.66    &  1.2    &  \nodata    & HCN, HNC\\[-1ex]
		&                IRAS18024-2119    &  18:05:25.4      &           -21:19:41     &8.83    & -0.03    &  0.1   &  \nodata    &  HNC\\[-1ex]
		&                IRAS18089-1732    & 18:11:51.3      &           -17:31:28     &12.89    &  0.49    & 13.0    &  3.6    &  HNC\\[-1ex]
		&                IRAS18102-1800    & 18:13:12.2      &           -17:59:35     &12.63    & -0.02    & 14.0    &  2.6    &  HNC\\[-1ex]
		&                IRAS18144-1723    & 18:17:24.4      &           -17:22:13     &13.66    & -0.60    &  4.3    &  \nodata    &  HNC\\[-1ex]
		&                IRAS18151-1208    & 18:17:57.1      &           -12:07:22     &18.34    &  1.77    &  3.0    &  \nodata    &  HCN, HNC\\[-1ex]
		&		  IRAS18162-1612	& 18:19:07.5     &		-16:11:21  & 14.89    &  -0.40 & 4.9	&	\nodata	& HNC\\[-1ex]
		&                IRAS18182-1433    & 18:21:07.9      &           -14:31:53     &16.58    & -0.05    & 11.8    &  4.5    & HCN, HNC\\[-1ex]
		&                IRAS18223-1243    & 18:25:10.9      &           -12:42:17     &18.66    & -0.06    & 12.4    &  3.7    &  HCN, HNC\\[-1ex]
		&                IRAS18264-1152    & 18:29:14.3      &           -11:50:26     &19.88    & -0.53    & 12.5    &  3.5    &   HCN, HNC\\[-1ex]
		&		  IRAS18290-0924	&18:31:44.8     &		-09:22:09	& 22.36    &   0.06 & 10.5	&	5.3	&	HNC\\[-1ex]
		&                IRAS18308-0841    & 18:33:31.9      &           -08:39:17     &23.20    &  0.00    & 10.7    &  4.9    &  HCN\\[-1ex]
		&		 IRAS18310-0825	& 18:33:47.2     &		-08:23:35	& 23.46    &  0.07 & 10.4	&	5.2	&	HNC\\[-1ex]
		&                IRAS18345-0641    & 18:37:16.8      &           -06:38:32     &25.41    &  0.10    &  9.5    &  \nodata    &  HNC\\[-1ex]
		&                IRAS18440-0148    & 18:46:36.3      &           -01:45:23     &30.82    &  0.27    &  8.3    &  \nodata    & HCN, HNC\\[-1ex]
		&		 IRAS18445-0222	& 18:47:10.8     &           -02:19:06	& 30.38    &  -0.11 & 9.4	&	5.3	&	HCN \\[-1ex]
		&		 IRAS18447-0229	& 18:47:23.7     &           -02:25:55	& 30.31    &  -0.21 & 8.2	&	6.6	&	HCN\\[-1ex]
		&                IRAS18470-0044    & 18:49:36.7      &           -00:41:05     &32.11    &  0.09    &  8.2    &  \nodata    & HCN, HNC\\[-1ex]
		&		 IRAS18488+0000	& 18:51:24.8     &           +00:04:19	& 32.99    &   0.04 & 8.9	&	5.4	&	HNC\\[-1ex]
		&                IRAS18511+0146    & 18:53:38.1      &           +01:50:27     &34.82    &  0.35    &  3.9    &  \nodata    &  HNC\\[-1ex]
		&		 IRAS18527+0301	& 18:55:16.5     &           +03:05:07	& 36.11    &  0.55 & 5.26	&	\nodata	& HCN\\[-1ex]
                &		 IRAS18530+0215    & 18:55:34.2      &           +02:19:08     &35.47    &  0.14    &  8.7    &  5.1    &  HCN, HNC\\[-1ex]
		&		 IRAS19012+0536	& 19:03:45.1     &           +05:40:40	& 39.39    &  -0.14 & 8.6	&	4.6	&HCN \\[-1ex]
		&		 IRAS19035+0641	& 19:06:01.1     &           +06:46:35	& 40.62    &  -0.14 & 2.2	&	\nodata	& HCN\\[-1ex]
		&                IRAS19220+1432    & 19:24:19.7      &           +14:38:03     &49.67    & -0.46    &  5.5    &  \nodata    & HNC\\[-1ex]
		&                IRAS19410+2336    & 19:43:11.4      &           +23:44:06     &59.78    &  0.06    &  6.4    &  2.1    & HCN  \\[-1ex]
		&                IRAS19411+2306    & 19:43:18.1      &           +23:13:59     &59.36    & -0.21    &  5.8    &  2.9    & HCN, HNC \\[-1ex]
		&                IRAS19413+2332	& 19:43:28.9     &           +23:40:04	& 59.76    &  -0.03 & 6.8	&	1.8	&	HCN \\[-1ex]
		&                IRAS20126+4104    & 20:14:26.0      &           +41:13:32     &78.12    &  3.63    &  1.7    &  \nodata    & HCN, HNC \\[-1ex]
		&                IRAS20216+4107	& 20:23:23.8     &           +41:17:40	& 79.12    &  2.28 & 1.7	&	\nodata	& HCN, HNC \\[-1ex]
		&                IRAS20293+3952	& 20:31:10.7     &           +40:03:10	& 78.98    &  0.36 & 2.0	&	1.3	& HNC \\[-1ex]
		&                IRAS20343+4129    & 20:36:07.1      &           +41:40:01     &80.83    &  0.57    &  1.4    &  \nodata    &  HCN, HNC\\[-1ex]
		&                IRAS22134+5834	& 22:15:09.1     &           +58:49:09	& 103.88    &  1.86 & 2.6	&	\nodata	& HCN, HNC\\[-1ex]
		&                IRAS22198+6336    & 22:21:27.6      &           +63:51:42     &107.30    &  5.64    &  1.3    &  \nodata    & HCN, HNC\\[-1ex]
		&                IRAS23033+5951    & 23:05:25.7      &           +60:08:08     &110.09    & -0.07    &  3.5    &  \nodata    & HCN, HNC\\[-1ex]	
		&                IRAS23140+6121	& 23:16:11.7     &           +61:37:45	& 111.87    &  0.82 & 6.44	&	\nodata	&HCN \\[-1ex]
UCHII	&                IRAS02232+6138    & 02:27:01.0      &           +61:52:14     &133.94    &  1.06    &  3.0    &	\nodata	& HCN, HNC\\[-1ex]
		&                IRAS02575+6017    & 03:01:32.3      &           +60:29:12     &138.30    &  1.56    &  3.8    &	\nodata	& HCN, HNC \\[-1ex]
		&                IRAS03035+5819    & 03:07:25.6      &           +58:30:52     &139.91    &  0.20    &  4.2    &	\nodata	& HCN\\[-1ex]
		&		  IRAS05393-0156	& 05:41:49.5	&		-01:55:17	& 206.56    &  -16.34	 & 0.5	&	\nodata	&  HNC\\[-1ex]
		&                IRAS06053-0622    & 06:07:46.6      &           -06:22:59     &213.70    &-12.60    & 10.8    &	\nodata	&  HCN \\[-1ex]
		&                IRAS06056+2131    &06:08:41.0      &           +21:31:01     &189.03    &  0.78    &  0.8    &	\nodata	&  HCN, HNC \\[-1ex]
		&                IRAS06058+2138    &06:08:54.1      &           +21:38:25     &188.95    &  0.89    &  2.2    &	\nodata	&HCN, HNC   \\[-1ex]
		&                IRAS06061+2151    &06:09:07.8      &           +21:50:39     &188.80    &  1.03    &  4.1    &	\nodata	& HCN \\[-1ex]
		&                IRAS06084-0611    & 06:10:51.0      &           -06:11:54     &213.88    &-11.84    &  1.0    &	\nodata	& HCN, HNC  \\[-1ex]
		&                IRAS06099+1800    &06:12:53.3      &           +17:59:22     &192.60    & -0.05    &  2.5    &	\nodata	&HCN, HNC \\[-1ex]
		&                IRAS17574-2403    & 18:00:30.4      &        -24:04:00     &5.89    & -0.39    &  2.0    &	\nodata	&  HNC\\[-1ex]
		&                IRAS17599-2148    &  18:03:00.4      &           -21:48:05     &8.14    &  0.23    &  4.2    &	\nodata	& HCN, HNC\\[-1ex]
		&                IRAS18032-2137    &  18:06:19.0      &        -21:37:32     &8.67    & -0.36    &  4.8    &	\nodata	&  HNC\\[-1ex]
		&                IRAS18075-1956    & 18:10:23.5      &         -19:56:15     &10.61    & -0.37    &  4.8    &	\nodata	& HCN, HNC  \\[-1ex]
		&                IRAS18100-1854    & 18:14:01.1      &         -18:53:24     &11.94    & -0.62    &  5.2    &	\nodata	& HNC \\[-1ex]
		&                IRAS18162-2048    &  18:19:11.9      &           -20:47:34     &10.84    &  -2.59    &  1.9    &	\nodata	& HCN, HNC \\[-1ex]
		&                IRAS18174-1612    & 18:20:24.8      &         -16:11:35     &15.03    & -0.68    &  2.1    &	\nodata	& HCN \\[-1ex]
 		&                IRAS18317-0757    & 18:34:24.9      &           -07:54:48     &23.95    &  0.15    &  6.0    &	\nodata	&  HCN\\[-1ex]
		&                IRAS18403-0417    & 18:42:58.2      &           -04:14:00     &28.20    & -0.05    &  9.1    &	\nodata	& HNC \\[-1ex]
		&                IRAS18434-0242	&18:46:03.9  &		-02:39:22 & 29.96  &  -0.03 &  7.4	&	\nodata	& HCN,HNC	\\[-1ex]
		&                IRAS18469-0132	&18:49:34.7	&	-01:29:08	&	31.40   &  -0.26	&	7.3	&	\nodata	& HCN \\[-1ex]
 		&                IRAS19095+0930    & 19:11:53.3      &         +09:35:46     &43.79    &  -0.13    &  9.0    &	\nodata	&HNC  \\[-1ex]
		&                IRAS20081+3122    &  20:10:09.1      &           +31:31:34     &69.54    & -0.98    &  3.0    &	\nodata	& HNC \\[-1ex]
		&                IRAS20255+3712    & 20:27:26.6      &           +37:22:48     &76.38    & -0.62    &  1.0    &	\nodata	&  HCN, HNC\\[-1ex]
		&                IRAS20178+4046    & 20:19:39.3      &           +40:56:30     &78.44    &  2.66    &  3.3    &	\nodata	&  HCN, HNC\\[-1ex]
		&                IRAS20350+4126    & 20:36:52.6      &           +41:36:32     &80.87    &  0.42    &  2.1    &	\nodata	& HCN, HNC \\[-1ex]
		&                IRAS22176+6303	& 22:19:18.2	&	+63:18:46 	& 106.80   &    5.31 &	0.9	&	\nodata	& HCN, HNC\\[-1ex]
		&                IRAS22543+6145    & 22:56:19.1      &           +62:01:57     &109.87    &  2.12    &  0.7    &	\nodata	& HCN, HNC\\[-1ex]
		&                IRAS23133+6050    &23:15:31.5      &           +61:07:09     &111.61    &  0.37    &  5.2    &	\nodata	&HCN \\[-1ex]		
		&                IRAS23138+5945    &23:16:04.8      &           +60:02:00     &111.28    & -0.66    &  2.5    &	\nodata	&HCN, HNC \\[-1ex]
\enddata
\label{sourceinform}
\tablecomments{Kinetic distances are quoted from \citet{simon06b}~(IRDCs), \citet{beuther02}, \citet{molinari96}~(HMPOs), and \citet{thompson06} and references therein~(UCHIIs). If the distance ambiguity is resolved, only far and no near distance is noted.}
\end{deluxetable} 

\begin{deluxetable}{llllllc}
\tablewidth{0pt}
\tabletypesize{\tiny}
\tablecolumns{          7}
\tablecaption{Derived HCN line parameters} 
\scriptsize
\tablehead{\colhead{Classification} & \colhead{Source name} & \colhead{$v_\textrm{thick}$} & \colhead{$v_\textrm{thin}$} & \colhead{$\Delta v_\textrm{thin}$} & \colhead{$\delta v$} & \colhead{Profile($\delta v$)}\\
\colhead{} & \colhead{} & \colhead{(km $\textrm{s}^{-1}$)} & \colhead{(km $\textrm{s}^{-1}$)} & \colhead{(km $\textrm{s}^{-1}$)} & \colhead{} & \colhead{}}
\startdata
qIRDCc&G028.37$\_$MM9&78.37 (0.041)&79.69 (0.217)&3.25 (0.509)&-0.41 (0.102)& Blue \\[-1ex]
&G035.39$\_$MM5&43.83 (0.054)&45.01 (0.156)&2.30 (0.377)&-0.51 (0.124)& Blue \\[-1ex]
aIRDCc&G022.35$\_$MM1$^{a}$&44.85 (0.190)&44.55 (0.162)&2.97 (0.404)&0.10 (0.119)& Neutral \\[-1ex]
&G030.97$\_$MM1&78.30 (0.035)&77.85 (0.068)&1.91 (0.146)&0.24 (0.057)& Neutral \\[-1ex]
&G031.97$\_$MM1&93.92 (0.211)&94.63 (0.079)&3.81 (0.272)&-0.19 (0.077)& Neutral \\[-1ex]
&G033.69$\_$MM4&104.46 (0.103)&106.00 (0.240)&3.66 (0.644)&-0.42 (0.120)& Blue \\[-1ex]
&G033.69$\_$MM5&105.49 (0.087)&105.13 (0.106)&1.57 (0.324)&0.23 (0.132)& Neutral \\[-1ex]
&G034.43$\_$MM4&54.86 (0.057)&57.47 (0.066)&3.46 (0.194)&-0.76 (0.055)& Blue \\[-1ex]
&G035.39$\_$MM7&45.02 (0.109)&45.56 (0.150)&1.83 (0.315)&-0.29 (0.150)& Blue \\[-1ex]
&G048.65$\_$MM1&33.82 (0.046)&34.22 (0.076)&0.87 (0.161)&-0.46 (0.165)& Blue \\[-1ex]
&G053.25$\_$MM4&24.57 (0.030)&24.12 (0.097)&1.09 (0.202)&0.41 (0.139)& Red \\[-1ex]
&G053.25$\_$MM6&23.45 (0.042)&23.31 (0.119)&1.26 (0.209)&0.11 (0.129)& Neutral \\[-1ex]
HMPO&IRAS00117+6412&-35.97 (0.079)&-36.03 (0.111)&0.92 (0.412)&0.06 (0.209)& Neutral \\[-1ex]
&IRAS05358+3543&-17.84 (0.025)&-17.31 (0.070)&1.95 (0.183)&-0.28 (0.055)& Blue \\[-1ex]
&IRAS05373+2349&2.09 (0.030)&2.29 (0.083)&1.56 (0.288)&-0.13 (0.076)& Neutral \\[-1ex]
&IRAS18151-1208&33.02 (0.014)&33.61 (0.100)&2.20 (0.236)&-0.27 (0.059)& Blue \\[-1ex]
&IRAS18182-1433&58.30 (0.211)&59.28 (0.116)&3.38 (0.270)&-0.29 (0.099)& Blue \\[-1ex]
&IRAS18223-1243&44.67 (0.021)&45.29 (0.079)&1.40 (0.217)&-0.44 (0.099)& Blue \\[-1ex]
&IRAS18264-1152&43.28 (0.034)&43.75 (0.054)&2.63 (0.158)&-0.18 (0.035)& Neutral \\[-1ex]
&IRAS18308-0841&77.72 (0.050)&76.81 (0.128)&2.87 (0.332)&0.32 (0.072)& Red \\[-1ex]
&IRAS18440-0148&97.53 (0.066)&98.03 (0.160)&2.29 (0.259)&-0.22 (0.102)& Neutral \\[-1ex]
&IRAS18445-0222&86.96 (0.045)&86.94 (0.259)&2.46 (0.672)&0.01 (0.123)& Neutral \\[-1ex]
&IRAS18447-0229&102.43 (0.133)&102.43 (0.134)&1.37 (0.291)&-0.00 (0.194)& Neutral \\[-1ex]
&IRAS18470-0044&97.27 (0.100)&96.30 (0.135)&2.99 (0.379)&0.32 (0.089)& Red \\[-1ex]
&IRAS18527+0301&74.94 (0.075)&75.75 (0.179)&2.42 (0.465)&-0.33 (0.123)& Blue \\[-1ex]
&IRAS18530+0215&77.07 (0.022)&77.08 (0.087)&2.78 (0.233)&-0.01 (0.039)& Neutral \\[-1ex]
&IRAS19012+0536&64.84 (0.111)&65.58 (0.151)&2.08 (0.385)&-0.36 (0.142)& Blue \\[-1ex]
&IRAS19035+0641&32.41 (0.078)&32.33 (0.184)&2.64 (0.502)&0.03 (0.099)& Neutral \\[-1ex]
&IRAS19410+2336&22.13 (0.016)&22.70 (0.116)&2.66 (0.287)&-0.22 (0.055)& Neutral \\[-1ex]
&IRAS19411+2306&29.31 (0.026)&29.34 (0.109)&1.62 (0.248)&-0.02 (0.084)& Neutral \\[-1ex]
&IRAS19413+2332&20.03 (0.024)&20.29 (0.132)&1.47 (0.256)&-0.18 (0.111)& Neutral \\[-1ex]
&IRAS20126+4104&-3.80 (0.025)&-3.65 (0.070)&2.55 (0.194)&-0.06 (0.037)& Neutral \\[-1ex]
&IRAS20216+4107&-1.65 (0.032)&-1.59 (0.094)&1.03 (0.218)&-0.07 (0.123)& Neutral \\[-1ex]
&IRAS20343+4129&11.48 (0.016)&11.48 (0.109)&2.38 (0.230)&0.00 (0.052)& Neutral \\[-1ex]
&IRAS22134+5834&-17.82 (0.076)&-18.39 (0.106)&1.69 (0.337)&0.33 (0.127)& Red \\[-1ex]
&IRAS22198+6336&-11.24 (0.029)&-11.05 (0.083)&1.32 (0.238)&-0.15 (0.089)& Neutral \\[-1ex]
&IRAS23033+5951&-53.18 (0.033)&-53.28 (0.143)&2.52 (0.318)&0.04 (0.070)& Neutral \\[-1ex]
&IRAS23140+6121&-53.18 (0.033)&-50.74 (0.244)&1.43 (0.496)&-1.71 (0.626)& Blue \\[-1ex]
UCHII&IRAS02232+6138&-47.78 (0.024)&-46.55 (0.035)&3.43 (0.092)&-0.36 (0.020)& Blue \\[-1ex]
&IRAS02575+6017&-38.13 (0.018)&-37.94 (0.060)&2.35 (0.150)&-0.08 (0.034)& Neutral \\[-1ex]
&IRAS03035+5819&-39.60 (0.024)&-39.57 (0.061)&1.69 (0.160)&-0.02 (0.050)& Neutral \\[-1ex]
&IRAS06053-0622&9.53 (0.029)&10.07 (0.084)&1.29 (0.204)&-0.42 (0.110)& Blue \\[-1ex]
&IRAS06056+2131&2.55 (0.007)&2.74 (0.057)&2.28 (0.137)&-0.08 (0.029)& Neutral \\[-1ex]
&IRAS06058+2138&3.36 (0.008)&3.35 (0.053)&2.69 (0.126)&0.01 (0.023)& Neutral \\[-1ex]
&IRAS06061+2151&-1.26 (0.015)&-0.30 (0.084)&2.58 (0.203)&-0.37 (0.048)& Blue \\[-1ex]
&IRAS06084-0611&11.65 (0.014)&11.09 (0.034)&2.88 (0.090)&0.20 (0.018)& Neutral \\[-1ex]
&IRAS06099+1800&7.33 (0.007)&7.35 (0.031)&2.30 (0.083)&-0.01 (0.017)& Neutral \\[-1ex]
&IRAS17599-2148&20.24 (0.023)&20.61 (0.265)&6.31 (0.746)&-0.06 (0.046)& Neutral \\[-1ex]
&IRAS18075-1956&-3.99 (0.113)&-2.95 (0.244)&4.81 (0.627)&-0.22 (0.079)& Neutral \\[-1ex]
&IRAS18162-2048&11.22 (0.043)&12.40 (0.062)&2.77 (0.154)&-0.43 (0.045)& Blue \\[-1ex]
&IRAS18174-1612&19.66 (0.014)&19.45 (0.059)&3.09 (0.162)&0.07 (0.024)& Neutral \\[-1ex]
&IRAS18317-0757&78.41 (0.050)&79.86 (0.067)&2.91 (0.167)&-0.50 (0.049)& Blue \\[-1ex]
&IRAS18434-0242&96.52 (0.027)&97.38 (0.051)&3.66 (0.137)&-0.23 (0.023)& Neutral \\[-1ex]
&IRAS18469-0132&87.37 (0.024)&88.09 (0.187)&3.04 (0.474)&-0.24 (0.079)& Neutral \\[-1ex]
&IRAS20255+3712&-1.52 (0.020)&-1.37 (0.084)&2.09 (0.255)&-0.07 (0.051)& Neutral \\[-1ex]
&IRAS20178+4046&1.03 (0.016)&1.03 (0.071)&1.64 (0.152)&0.00 (0.053)& Neutral \\[-1ex]
&IRAS20350+4126&-3.88 (0.035)&-2.85 (0.089)&2.33 (0.229)&-0.44 (0.069)& Blue \\[-1ex]
&IRAS22176+6303&-6.80 (0.007)&-6.77 (0.030)&2.63 (0.071)&-0.01 (0.014)& Neutral \\[-1ex]
&IRAS22543+6145&-12.41 (0.018)&-10.40 (0.056)&3.01 (0.144)&-0.67 (0.040)& Blue \\[-1ex]
&IRAS23133+6050&-56.48 (0.020)&-56.16 (0.055)&2.49 (0.148)&-0.13 (0.031)& Neutral \\[-1ex]
&IRAS23138+5945&-44.56 (0.055)&-44.64 (0.145)&3.33 (0.339)&0.02 (0.060)& Neutral \\[-1ex]
\enddata
\label{hcnlinepara}
\tablecomments{$^{a}$ -- The $F$=0--1 hyperfine component is adopted as a standard for \deltav\ calculation.}
\end{deluxetable}

\begin{deluxetable}{llllllcccc}
\tablewidth{0pt}
\tabletypesize{\tiny}
\tablecolumns{          10}
\tablecaption{Derived HNC line parameters} \scriptsize
\tablehead{\colhead{Classification} & \colhead{Source name} & \colhead{$v_\textrm{thick}$} & \colhead{$v_\textrm{thin}$} & \colhead{$\Delta v_\textrm{thin}$} & \colhead{$\delta v$} & \colhead{Profile($\delta v$)} & 
\colhead{$\frac{T(B)}{T(R)}$} & \colhead{Profile($\frac{T(B)}{T(R)}$)} & \colhead{$v_\textrm{dip}$}\\
\colhead{} & \colhead{} & \colhead{(km $\textrm{s}^{-1}$)} & \colhead{(km $\textrm{s}^{-1}$)} & \colhead{(km $\textrm{s}^{-1}$)} & \colhead{} & \colhead{} & \colhead{} & \colhead{} & \colhead{(km $\textrm{s}^{-1}$)}}
\startdata
qIRDCc&G028.37$\_$MM9&78.48 (0.028)&80.16 (0.062)&2.63 (0.144)&-0.64 (0.049)& Blue   &  5.66  &  Blue  &  80.82 \\[-1ex]
&G031.97$\_$MM9&96.06 (0.051)&96.75 (0.202)&5.39 (0.456)&-0.13 (0.048)& Neutral   &  1.21  &  Neutral  &  97.27 \\[-1ex]
&G035.39$\_$MM5&44.12 (0.029)&45.51 (0.059)&1.64 (0.165)&-0.85 (0.101)& Blue  &  \nodata  &  \nodata  &  \nodata  \\[-1ex]
aIRDCc&G015.31$\_$MM3&31.27 (0.105)&31.04 (0.057)&0.99 (0.116)&0.23 (0.166)& Neutral  &  \nodata  &  \nodata  &  \nodata  \\[-1ex]
&G023.60$\_$MM1&106.75 (0.044)&106.74 (0.137)&4.35 (0.319)&0.00 (0.042)& Neutral   &  0.94  &  Neutral  &  106.25 \\[-1ex]
&G023.60$\_$MM6&53.10 (0.040)&53.23 (0.081)&1.77 (0.182)&-0.08 (0.069)& Neutral  &  \nodata  &  \nodata  &  \nodata  \\[-1ex]
&G024.60$\_$MM1&54.11 (0.035)&53.11 (0.091)&1.88 (0.196)&0.53 (0.087)& Red  &  \nodata  &  \nodata  &  \nodata  \\[-1ex]
&G028.37$\_$MM4&78.50 (0.023)&79.31 (0.066)&3.17 (0.157)&-0.41 (0.035)& Blue  &  \nodata  &  \nodata  &  \nodata  \\[-1ex]
&G028.37$\_$MM6&78.32 (0.054)&80.29 (0.043)&2.37 (0.105)&-0.83 (0.055)& Blue   &  3.26  &  Blue  &  81.44 \\[-1ex]
&G030.97$\_$MM1&77.95 (0.018)&77.95 (0.075)&2.47 (0.197)&-0.00 (0.038)& Neutral  &  \nodata  &  \nodata  &  \nodata  \\[-1ex]
&G031.97$\_$MM1&94.34 (0.026)&95.42 (0.058)&3.51 (0.140)&-0.31 (0.027)& Blue  &  \nodata  &  \nodata  &  \nodata  \\[-1ex]
&G031.97$\_$MM8&94.25 (0.065)&94.70 (0.079)&2.46 (0.245)&-0.18 (0.061)& Neutral   &  2.62  &  Blue  &  97.89 \\[-1ex]
&G033.69$\_$MM4&105.78 (0.048)&106.11 (0.071)&3.35 (0.180)&-0.10 (0.036)& Neutral  &  \nodata  &  \nodata  &  \nodata  \\[-1ex]
&G033.69$\_$MM5&105.20 (0.035)&105.25 (0.073)&1.97 (0.193)&-0.02 (0.055)& Neutral  &  \nodata  &  \nodata  &  \nodata  \\[-1ex]
&G034.43$\_$MM1&58.14 (0.024)&57.79 (0.052)&2.85 (0.116)&0.12 (0.027)& Neutral   &  0.56  &  Red  &  57.26 \\[-1ex]
&G034.43$\_$MM3&59.48 (0.029)&59.41 (0.039)&2.15 (0.103)&0.03 (0.032)& Neutral  &  \nodata  &  \nodata  &  \nodata  \\[-1ex]
&G034.43$\_$MM4&56.75 (0.035)&57.76 (0.039)&2.47 (0.098)&-0.41 (0.034)& Blue  &  \nodata  &  \nodata  &  \nodata  \\[-1ex]
&G035.59$\_$MM3&44.40 (0.036)&44.76 (0.065)&1.34 (0.137)&-0.27 (0.080)& Blue  &  \nodata  &  \nodata  &  \nodata  \\[-1ex]
&G038.95$\_$MM3&42.58 (0.040)&42.27 (0.094)&2.20 (0.249)&0.14 (0.063)& Neutral  &  \nodata  &  \nodata  &  \nodata  \\[-1ex]
&G038.95$\_$MM4&42.03 (0.040)&42.30 (0.107)&1.40 (0.307)&-0.19 (0.113)& Neutral  &  \nodata  &  \nodata  &  \nodata  \\[-1ex]
&G048.65$\_$MM1&33.78 (0.030)&34.03 (0.120)&1.65 (0.257)&-0.15 (0.094)& Neutral  &  \nodata  &  \nodata  &  \nodata  \\[-1ex]
&G048.65$\_$MM2&33.62 (0.032)&33.85 (0.106)&1.07 (0.209)&-0.22 (0.136)& Neutral  &  \nodata  &  \nodata  &  \nodata  \\[-1ex]
&G053.25$\_$MM4&24.48 (0.012)&24.43 (0.040)&1.19 (0.094)&0.04 (0.044)& Neutral  &  \nodata  &  \nodata  &  \nodata  \\[-1ex]
&G053.25$\_$MM6&23.55 (0.019)&23.74 (0.089)&1.45 (0.178)&-0.13 (0.077)& Neutral  &  \nodata  &  \nodata  &  \nodata  \\[-1ex]
&G053.31$\_$MM2&25.03 (0.029)&25.57 (0.086)&1.31 (0.216)&-0.41 (0.111)& Blue  &  \nodata  &  \nodata  &  \nodata  \\[-1ex]
HMPO&IRAS05358+3543&-17.64 (0.013)&-17.40 (0.076)&1.74 (0.201)&-0.14 (0.054)& Neutral  &  \nodata  &  \nodata  &  \nodata  \\[-1ex]
&IRAS05373+2349&2.26 (0.016)&2.42 (0.072)&1.57 (0.218)&-0.10 (0.058)& Neutral  &  \nodata  &  \nodata  &  \nodata  \\[-1ex]
&IRAS18024-2119&0.93 (0.117)&0.60 (0.063)&2.11 (0.153)&0.15 (0.086)& Neutral  &  \nodata  &  \nodata  &  \nodata  \\[-1ex]
&IRAS18089-1732&34.34 (0.069)&32.78 (0.080)&3.39 (0.170)&0.46 (0.050)& Red   &  0.47  &  Red  &  33.36 \\[-1ex]
&IRAS18102-1800&21.67 (0.058)&21.41 (0.086)&2.07 (0.196)&0.13 (0.071)& Neutral  &  \nodata  &  \nodata  &  \nodata  \\[-1ex]
&IRAS18144-1723&47.86 (0.040)&47.61 (0.076)&3.29 (0.180)&0.08 (0.035)& Neutral   &  0.77  &  Red  &  47.78 \\[-1ex]
&IRAS18151-1208&33.10 (0.020)&33.37 (0.073)&1.89 (0.180)&-0.14 (0.051)& Neutral  &  \nodata  &  \nodata  &  \nodata  \\[-1ex]
&IRAS18162-1612&61.86 (0.028)&61.86 (0.092)&1.77 (0.228)&-0.00 (0.068)& Neutral  &  \nodata  &  \nodata  &  \nodata  \\[-1ex]
&IRAS18182-1433&58.85 (0.031)&59.66 (0.112)&3.04 (0.258)&-0.26 (0.052)& Blue  &  \nodata  &  \nodata  &  \nodata  \\[-1ex]
&IRAS18223-1243&45.33 (0.025)&45.29 (0.067)&1.85 (0.156)&0.03 (0.050)& Neutral  &  \nodata  &  \nodata  &  \nodata  \\[-1ex]
&IRAS18264-1152&43.99 (0.017)&43.83 (0.076)&2.41 (0.183)&0.07 (0.039)& Neutral  &  \nodata  &  \nodata  &  \nodata  \\[-1ex]
&IRAS18290-0924&84.29 (0.064)&84.47 (0.111)&1.98 (0.267)&-0.09 (0.089)& Neutral  &  \nodata  &  \nodata  &  \nodata  \\[-1ex]
&IRAS18310-0825&84.46 (0.076)&84.67 (0.089)&1.83 (0.222)&-0.12 (0.091)& Neutral  &  \nodata  &  \nodata  &  \nodata  \\[-1ex]
&IRAS18345-0641&95.46 (0.082)&95.48 (0.063)&1.79 (0.152)&-0.01 (0.081)& Neutral  &  \nodata  &  \nodata  &  \nodata  \\[-1ex]
&IRAS18440-0148&97.69 (0.029)&97.67 (0.153)&1.47 (0.433)&0.02 (0.124)& Neutral  &  \nodata  &  \nodata  &  \nodata  \\[-1ex]
&IRAS18470-0044&96.49 (0.067)&96.31 (0.162)&2.52 (0.462)&0.07 (0.092)& Neutral   &  0.59  &  Red  &  95.71 \\[-1ex]
&IRAS18488+0000&82.78 (0.141)&83.36 (0.182)&2.95 (0.439)&-0.20 (0.113)& Neutral  &  \nodata  &  \nodata  &  \nodata  \\[-1ex]
&IRAS18511+0146&57.10 (0.038)&56.94 (0.061)&1.77 (0.145)&0.09 (0.056)& Neutral  &  \nodata  &  \nodata  &  \nodata  \\[-1ex]
&IRAS18530+0215&77.25 (0.016)&77.30 (0.100)&2.73 (0.219)&-0.02 (0.042)& Neutral  &  \nodata  &  \nodata  &  \nodata  \\[-1ex]
&IRAS19220+1432&69.27 (0.104)&69.78 (0.298)&3.11 (0.516)&-0.16 (0.132)& Neutral  &  \nodata  &  \nodata  &  \nodata  \\[-1ex]
&IRAS19411+2306&29.35 (0.019)&29.24 (0.088)&1.60 (0.233)&0.12 (0.110)& Neutral  &  \nodata  &  \nodata  &  \nodata  \\[-1ex]
&IRAS20126+4104&-3.45 (0.015)&-3.95 (0.048)&1.95 (0.110)&0.26 (0.035)& Red  &  \nodata  &  \nodata  &  \nodata  \\[-1ex]
&IRAS20216+4107&-1.57 (0.025)&-1.63 (0.054)&0.97 (0.118)&0.06 (0.082)& Neutral  &  \nodata  &  \nodata  &  \nodata  \\[-1ex]
&IRAS20293+3952&6.06 (0.026)&5.96 (0.117)&2.32 (0.281)&0.04 (0.062)& Neutral  &  \nodata  &  \nodata  &  \nodata  \\[-1ex]
&IRAS20343+4129&11.15 (0.022)&11.73 (0.123)&2.29 (0.261)&-0.25 (0.070)& Blue  &  \nodata  &  \nodata  &  \nodata  \\[-1ex]
&IRAS22134+5834&-18.24 (0.064)&-18.80 (0.142)&1.54 (0.260)&0.37 (0.147)& Red   &  0.94  &  Neutral  &  -18.40 \\[-1ex]
&IRAS22198+6336&-11.12 (0.061)&-10.97 (0.065)&1.25 (0.164)&-0.12 (0.102)& Neutral  &  \nodata  &  \nodata  &  \nodata  \\[-1ex]
&IRAS23033+5951&-52.79 (0.034)&-53.69 (0.185)&3.13 (0.450)&0.29 (0.081)& Red  &  \nodata  &  \nodata  &  \nodata  \\[-1ex]
UCHII&IRAS02232+6138&-46.88 (0.033)&-46.35 (0.077)&2.90 (0.178)&-0.18 (0.040)& Neutral  &  \nodata  &  \nodata  &  \nodata  \\[-1ex]
&IRAS02575+6017&-38.24 (0.019)&-37.97 (0.067)&1.37 (0.145)&-0.20 (0.066)& Neutral  &  \nodata  &  \nodata  &  \nodata  \\[-1ex]
&IRAS05393-0156&10.34 (0.046)&9.47 (0.074)&0.57 (0.148)&1.52 (0.447)& Red   &  1.09  &  Neutral  &  10.57 \\[-1ex]
&IRAS06056+2131&2.67 (0.010)&2.83 (0.107)&2.58 (0.277)&-0.06 (0.046)& Neutral  &  \nodata  &  \nodata  &  \nodata  \\[-1ex]
&IRAS06058+2138&3.31 (0.010)&3.40 (0.075)&2.00 (0.182)&-0.04 (0.043)& Neutral  &  \nodata  &  \nodata  &  \nodata  \\[-1ex]
&IRAS06084-0611&11.69 (0.020)&11.31 (0.077)&2.25 (0.185)&0.17 (0.045)& Neutral  &  \nodata  &  \nodata  &  \nodata  \\[-1ex]
&IRAS06099+1800&7.32 (0.010)&7.32 (0.082)&1.79 (0.205)&-0.00 (0.052)& Neutral  &  \nodata  &  \nodata  &  \nodata  \\[-1ex]
&IRAS17574-2403&9.01 (0.017)&8.95 (0.032)&3.78 (0.087)&0.02 (0.013)& Neutral  &  \nodata  &  \nodata  &  \nodata  \\[-1ex]
&IRAS17599-2148&20.10 (0.031)&18.87 (0.106)&4.15 (0.241)&0.30 (0.037)& Red   &  0.38  &  Red  &  18.00 \\[-1ex]
&IRAS18032-2137&33.50 (0.047)&35.10 (0.041)&4.48 (0.110)&-0.36 (0.021)& Blue   &  3.32  &  Blue  &  37.74 \\[-1ex]
&IRAS18075-1956&-3.13 (0.029)&-2.00 (0.193)&2.75 (0.472)&-0.41 (0.107)& Blue  &  \nodata  &  \nodata  &  \nodata  \\[-1ex]
&IRAS18162-2048&12.36 (0.024)&12.46 (0.130)&2.37 (0.262)&-0.04 (0.065)& Neutral   &  1.30  &  Blue  &  12.51 \\[-1ex]
&IRAS18100-1854&39.96 (0.037)&38.37 (0.051)&3.95 (0.112)&0.40 (0.025)& Red   &  0.41  &  Red  &  36.99 \\[-1ex]
&IRAS18403-0417&96.26 (0.053)&95.74 (0.071)&4.44 (0.186)&0.12 (0.028)& Neutral   &  1.62  &  Blue  &  97.56 \\[-1ex]
&IRAS18434-0242&96.85 (0.015)&97.56 (0.055)&3.02 (0.137)&-0.24 (0.026)& Neutral  &  \nodata  &  \nodata  &  \nodata  \\[-1ex]
&IRAS19095+0930&44.12 (0.062)&44.12 (0.247)&4.77 (0.610)&0.00 (0.065)& Neutral  &  \nodata  &  \nodata  &  \nodata  \\[-1ex]
&IRAS20081+3122&12.52 (0.022)&11.74 (0.068)&3.91 (0.168)&0.20 (0.025)& Neutral  &  \nodata  &  \nodata  &  \nodata  \\[-1ex]
&IRAS20255+3712&-1.41 (0.019)&-1.49 (0.074)&1.92 (0.215)&0.04 (0.049)& Neutral  &  \nodata  &  \nodata  &  \nodata  \\[-1ex]
&IRAS20178+4046&0.98 (0.012)&1.07 (0.097)&1.87 (0.216)&-0.05 (0.058)& Neutral  &  \nodata  &  \nodata  &  \nodata  \\[-1ex]
&IRAS20350+4126&-3.01 (0.035)&-2.46 (0.110)&2.27 (0.282)&-0.24 (0.071)& Neutral  &  \nodata  &  \nodata  &  \nodata  \\[-1ex]
&IRAS22176+6303&-6.84 (0.008)&-6.83 (0.082)&2.27 (0.184)&-0.01 (0.040)& Neutral  &  \nodata  &  \nodata  &  \nodata  \\[-1ex]
&IRAS22543+6145&-11.38 (0.024)&-10.61 (0.117)&2.80 (0.268)&-0.27 (0.057)& Blue   &  1.78  &  Blue  &  -10.12 \\[-1ex]
&IRAS23138+5945&-44.49 (0.031)&-44.44 (0.155)&2.12 (0.381)&-0.03 (0.088)& Neutral  &  \nodata  &  \nodata  &  \nodata  \\[-1ex]
\enddata
\label{hnclinepara}
\end{deluxetable}

\begin{deluxetable}{lcccccc}
\tablewidth{0pt}
\tabletypesize{\footnotesize}
\tablecolumns{7}
\tablecaption{Blue excess~($E$) statistics}
\scriptsize
\tablehead{
\colhead{Evolutionary stage} 	& \colhead{Inflow Tracer}	 & \colhead{$N_\textrm{Blue}$}	& \colhead{$N_\textrm{Red}$}	& \colhead{$N_\textrm{Total}$}	& \colhead{$E$} & \colhead{$P$}
}
\startdata
IRDCs	& HCN	& 6	& 1 & 12	& 0.42	& 0.062\\
		& HNC	& 8 	& 1 & 25 	& 0.28	& 0.019\\
HMPOs	& HCN	& 7	& 3 & 26	& 0.15	& 0.172\\
		& HNC	& 2	& 4 & 28	& -0.07	& 0.891\\
UCHIIs	& HCN	& 7	& 0 & 23	& 0.30	& 0.008\\
		& HNC	& 3	& 3 & 23	& 0.00	& 0.500\\
\enddata
\label{statresult}

\end{deluxetable} 

\end{document}